\documentclass[conference]{IEEEtran}
\IEEEoverridecommandlockouts
\usepackage{listings}
\usepackage[linesnumbered,ruled]{algorithm2e}
\SetKwInput{KwInput}{Input}                
\SetKwInput{KwOutput}{Output}              
\usepackage{amsmath,amssymb,amsfonts}
\usepackage{algorithmic}
\usepackage{graphicx}
\usepackage{tabularx}
\usepackage{textcomp}
\usepackage{xcolor}
\usepackage{float}
\usepackage{subfig}
\usepackage{amssymb}
\usepackage{pifont}
\newcommand{\cmark}{\ding{51}}%
\newcommand{\xmark}{\ding{55}}%
\usepackage[T1]{fontenc}
\usepackage[noadjust]{cite}
\def\BibTeX{{\rm B\kern-.05em{\sc i\kern-.025em b}\kern-.08em
    T\kern-.1667em\lower.7ex\hbox{E}\kern-.125emX}}

\usepackage{listings}
\definecolor{comment}{rgb}{0,.6,0}
\definecolor{keyword}{rgb}{.63,0,.42}
\definecolor{kw2}{rgb}{.50,.50,.15}
\definecolor{kw3}{rgb}{.42,.42,.63}
\definecolor{string}{rgb}{1,0,0}

\definecolor{mGreen}{rgb}{0,0.6,0}
\definecolor{mGray}{rgb}{0.5,0.5,0.5}
\definecolor{mPurple}{rgb}{0.58,0,0.82}
\definecolor{backgroundColour}{rgb}{0.95,0.95,0.92}

\lstdefinestyle{VerilogStyle}{
    backgroundcolor=\color{backgroundColour}
    language=Verilog,
    commentstyle=\color{comment},
    stringstyle=\color{string},
    keywordstyle=\bfseries\color{keyword},
    basicstyle=\footnotesize\ttfamily,
    numbers=none,
    numberstyle=\tiny,
    numbersep=5pt,
    frame=lines,
    breaklines=true,
    prebreak=\raisebox{0ex}[0ex][0ex]{\ensuremath{\hookleftarrow}},
    showstringspaces=false,
    upquote=true,
    tabsize=2,
    xleftmargin=0em,
    xrightmargin=0em,
}

\definecolor{mycolor}{rgb}{0.90,0.95,0.90}
\lstdefinestyle{code}{
    backgroundcolor=\color{mycolor},   
    commentstyle=\color{codegreen},
    keywordstyle=\color{magenta},
    numberstyle=\tiny\color{codegray},
    stringstyle=\color{codepurple},
    basicstyle=\ttfamily\footnotesize,
    breakatwhitespace=false,         
    breaklines=true,                 
    captionpos=b,                    
    keepspaces=true,   
    numbers=none,                      
    showspaces=false,                
    showstringspaces=false,
    showtabs=false,                  
    tabsize=2
}
\usepackage{multirow}

\begin{document}

\title{DiSPEL: Distributed Security Policy Enforcement for Bus-based SoC\\
}

\author{\IEEEauthorblockN{Sudipta Paria}
\IEEEauthorblockA{\textit{Electrical and Computer Engineering} \\
\textit{University of Florida}\\
Gainesville, USA \\
sudiptaparia@ufl.edu}
\and
\IEEEauthorblockN{Swarup Bhunia}
\IEEEauthorblockA{\textit{Electrical and Computer Engineering} \\
\textit{University of Florida}\\
Gainesville, USA \\
swarup@ece.ufl.edu}

}

\maketitle

\begin{abstract}
The current zero trust model adopted in System-on-Chip (SoC) design is vulnerable to various malicious entities, and modern SoC designs must incorporate various security policies to protect sensitive assets from unauthorized access. These policies involve complex interactions between multiple IP blocks, which poses challenges for SoC designers and security experts when implementing these policies and for system validators when ensuring compliance. Difficulties arise when upgrading policies, reusing IPs for systems targeting different security requirements, and the subsequent increase in design time and time-to-market. This paper proposes a generic and flexible framework, called DiSPEL, for enforcing security policies defined by the user represented in a formal way for any bus-based SoC design. It employs a distributed deployment strategy while ensuring trusted bus operations despite the presence of untrusted IPs. It relies on incorporating a dedicated, centralized module capable of implementing diverse security policies involving bus-level interactions while generating the necessary logic and appending in the bus-level wrapper for IP-level policies. The proposed architecture is generic and independent of specific security policy types supporting both synthesizable and non-synthesizable solutions. The experimental results demonstrate its effectiveness and correctness in enforcing the security requirements and viability due to low overhead in terms of area, delay, and power consumption tested on open-source standard SoC benchmarks.
\end{abstract}


\section{Introduction}

Computing systems in modern times are commonly developed as System-on-Chip (SoC) designs, seamlessly integrating numerous functionalities into a single integrated circuit.  These SoC designs comprise various design modules, also known as Intellectual Properties (IPs), that collaborate through on-chip communication fabrics to achieve the desired system functionality. Designing a secure SoC becomes more complex due to the presence of security vulnerabilities introduced at various stages of the design flow. In a typical SoC design, secure assets are distributed among various IPs and require protection from unauthorized access. These assets include crucial components such as cryptographic keys, Digital Rights Management (DRM) keys, programmable fuses, on-chip debug instrumentation, defeature bits, and more. In order to adequately safeguard the secure assets of an SoC from these threats, it is imperative to incorporate design-time considerations as a necessary measure to prevent potential attacks or felicitate detection and recovery in the event of an attack. In modern SoC designs, secure assets are distributed across various IPs, requiring protection against unauthorized access which can lead to severe consequences such as identity theft, leakage of sensitive data, and substantial financial losses etc. To safeguard these secure assets, the implementation of security policies is essential, as they outline the authentication, access, and protection requirements for various assets within the design. Typically, these policies reflect confidentiality, integrity, and availability requirements at a high level. By instantiating these requirements for specific assets, security policies provide actionable specifications for SoC system architects and designers, guiding them in implementing appropriate protection and mitigation strategies. Security policies ensure that access and updates to secure assets adhere to the CIA (Confidentiality, Integrity, Availability) properties. 

Here are two representative examples of security policies for a typical System-on-Chip (SoC). It is important to note that these policies are purely for demonstrative purposes and do not represent the comprehensive set of security policies for any particular company or design. 

\begin{itemize}
    \item \textit{Example 1: Key authentication} \\
    The crypto engine will only provide actual keys in response to key access requests from other IPs if the requesting IP has been authenticated in test mode. 
    
    \item \textit{Example 2: Boot confidentiality} \\
    At boot time, all internal registers of the crypto engine are inaccessible to any IP.
\end{itemize}

\begin{figure}[!ht]
\centering
\includegraphics[scale=0.5]{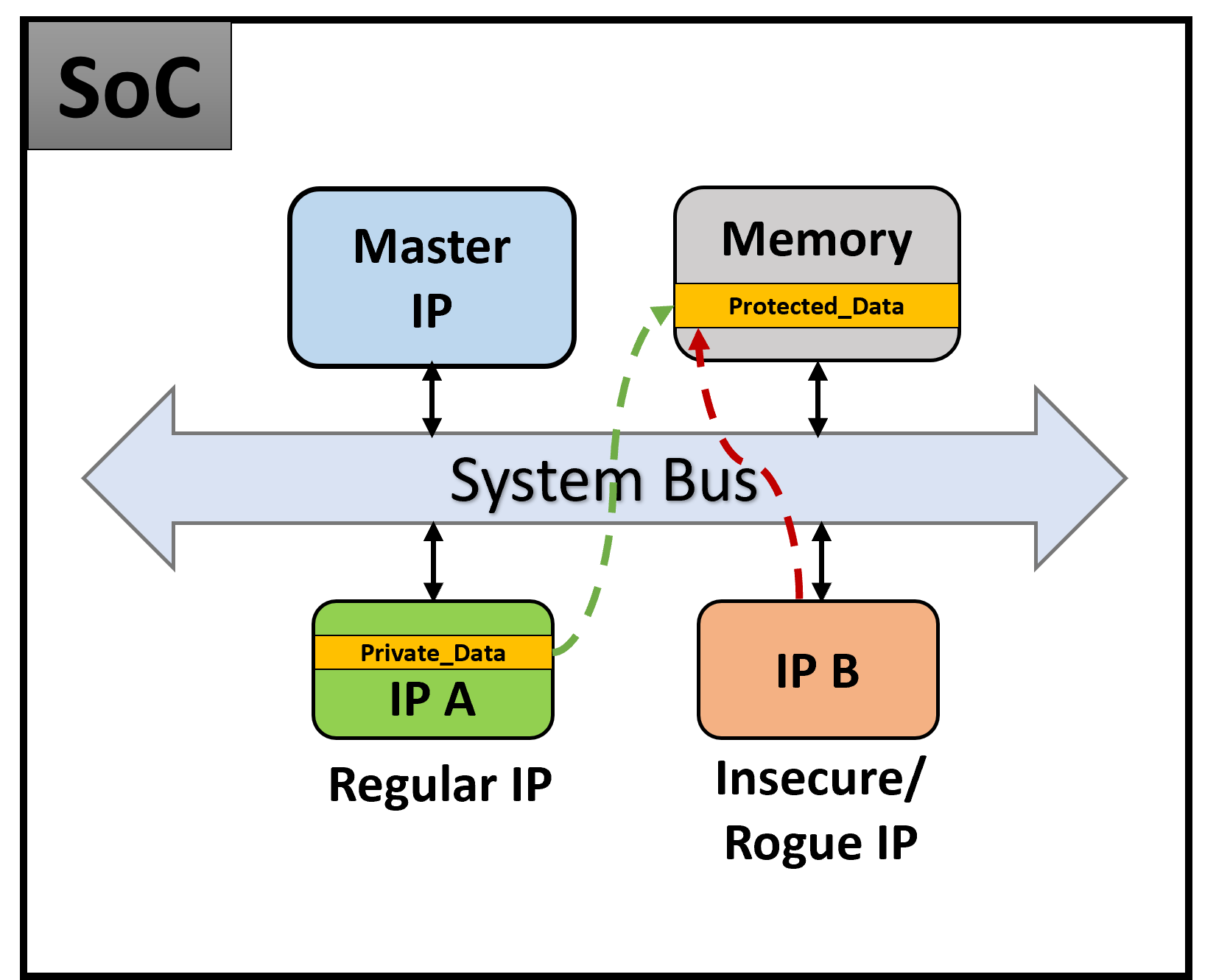}
\caption{Representation of a Security Threat: Accessing Private Data by Insecure/Rogue IP in a Generalized bus-based SoC}
\label{fig:threat_model}
\end{figure}

In Fig. \ref{fig:threat_model}, a comprehensive SoC model is depicted, featuring several essential components connected through a common bus, including a master IP, a memory IP, a regular IP containing sensitive data to be protected, and an insecure or rogue IP attempting to gain unauthorized access to the protected data. This illustration highlights a scenario analogous to {\em Example 1}, where the memory IP requests sensitive data from IP A while IP B attempts to access the protected data during transit. This scenario highlights the need for implementing a robust security mechanism to safeguard sensitive data and uphold the confidentiality, integrity, and availability of the overall design.

Designing authentication mechanisms to ensure policy enforcement has become increasingly challenging due to a multitude of complex policies and protection requirements across various classes of potential adversaries. Current methodologies heavily depend on manual efforts with creativity and observation by security experts. The typical process involves following two steps after designing a baseline architecture definition and then refining it in an iterative manner:
\begin{itemize}
    \item Employ threat modeling to recognize potential threats to the current architecture.
    \item Enhance the architecture by integrating mitigation strategies to address the identified threats.
\end{itemize}

The responsibility of Security Architects or Experts includes identifying three crucial elements for each asset: (1) identifying authorized users with access to the asset, (2) defining the permissible types of access according to the policies, and (3) determining the specific stages during the execution of the system or development/deployment life cycle when access requests can be approved or denied. However, this procedure can be exceedingly intricate and time-consuming, mainly because of the significant number of assets present in a typical SoC design. These assets can either have static definitions or be dynamically generated at different IPs during the execution of the system, which adds to the overall complexity. As SoC designs become increasingly intricate, integrating Design-for-Security (DfS) mechanisms has become a significant challenge. The main obstacles are as follows: (1) addressing threats at the architecture level along with IP level issues, (2) incurring acceptable additional overhead while upholding the functional correctness of IPs, (3) avoiding design or test conflicts due to multiple DfS approaches, (4) achieving comprehensive protection when using insecure third-party IPs.

\begin{figure*}[!ht]
\centering
\includegraphics[scale=0.5]{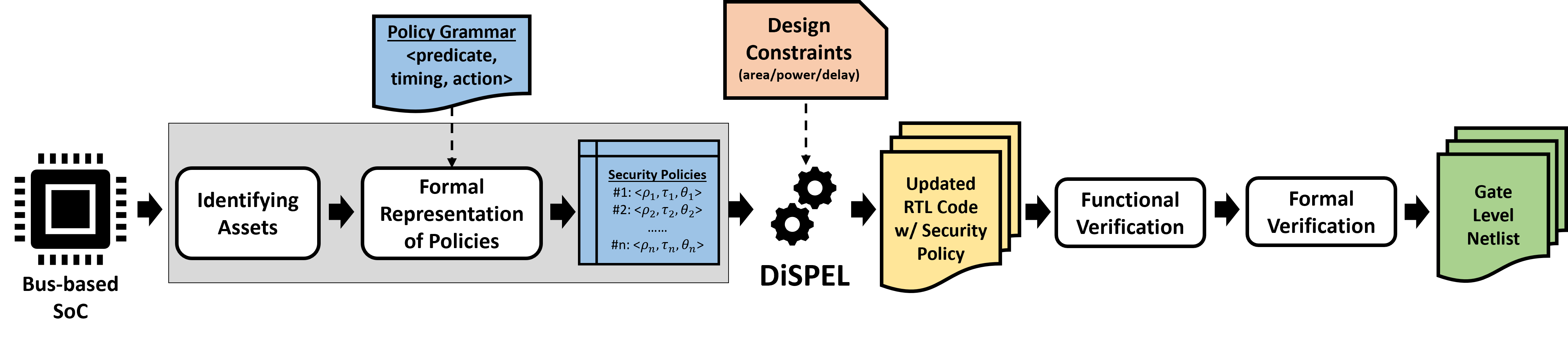}
\caption{DiSPEL: Proposed workflow for enforcing Security Policies represented in 3-tuple format for any bus-based SoC}
\label{fig:proposed_arch}
\end{figure*}

The existing proposed solutions, such as IIPS (\cite{b17}), E-IIPS (\cite{b12}), RSPE (\cite{b26}), MSIPS (\cite{b23}) offer policy enforcement through smart wrappers and infrastructure IPs in bus-based SoC primarily handling the architecture-level or IP core-level threats separately at runtime. However, they do not cater to diverse customizable security needs as defined by the user using a high-level language and lack flexibility and extensibility due to significant manual efforts involved in the workflow. Formalizing the security requirements and incorporating automation flow helps eliminate the dependence on security experts, reduces manual efforts, enhances response time, and enables a more proactive and effective approach to threat detection and mitigation efforts. Additionally, automation enables scalable and streamlined verification of the security policies with improved verification efficiency than manual testing efforts at different stages of the SoC development cycle in the zero-trust environment. Deb Nath et al. (\cite{b27}) proposed an architecture to facilitate the formal verification of security policies enforced in an SoC but has limitations in terms of generalization and applicability to a wider range of security scenarios for any bus-based SoC implementation. 

In this paper, we propose an automated framework named DiSPEL (Distributed Security Policies Enforcement Logic) for addressing the bus-level  security requirements involving multiple IPs and IP-level threats and incorporating low-overhead preventive measures through security policies. Fig. \ref{fig:proposed_arch} presents the overall workflow for incorporating security policies represented using a 3-tuple format in a bus-based SoC. The automated tool enforces the policies through a centralized module in a distributed manner ensuring minimal overhead while validating the correctness and effectiveness of the policies through simulation and formal verification. 

The major contributions of this work are as follows:
\begin{itemize}
    \item An automated framework for enforcing system-level security requirements through synthesizable security policies represented in a formal way for any bus-based SoC with minimal overhead. 
    \item The enforcement of bus-level security policies involving multiple IPs through a centralized policy module, along with the support for IP-level policies through bus-level wrappers.
    \item The convenient option of configurable security specifications presented in a high-level format through security policies by the end user that can be extended for different bus-based SoC configurations under any specific threat model and trust assumptions.
    \item The experimental results provide insights into the effectiveness and capability of the proposed flow in efficiently achieving SoC security, while the simulation \& formal verification results affirm the correctness of the overall framework. 
\end{itemize}

The remainder of the paper is organized as follows: Section II provides the background and an overview of the related work. Section III presents the methodology of the overall framework. The experimental results are described in Section IV. Finally,  we conclude this paper in Section V.

\section{Background \& Related Work}

\subsection{Security Policy}

Security Policies refer to a set of guidelines, rules, or specifications that are designed to ensure the protection and integrity of assets within the SoC design. Security policies serve a significant part in determining the access, protection, and authentication requirements of different assets within the design. These policies typically represent instances of CIA (Confidentiality, Integrity, and Availability) needs at a higher level. Security policies are put in place to prevent unauthorized access, detect and mitigate potential security threats, and maintain the overall security posture of the system. They serve as actionable instructions for the SoC system architect and designer to implement security measures and countermeasures effectively, ensuring the confidentiality, integrity, and availability of critical assets within the SoC. Security policies in modern SoC designs are often complex and developed in an ad-hoc manner due to their inherent intricacies and undergo refinement and modification throughout the system development process based on security requirements and product needs. 

\begin{figure}[!ht]
\centering
\includegraphics[scale=0.5]{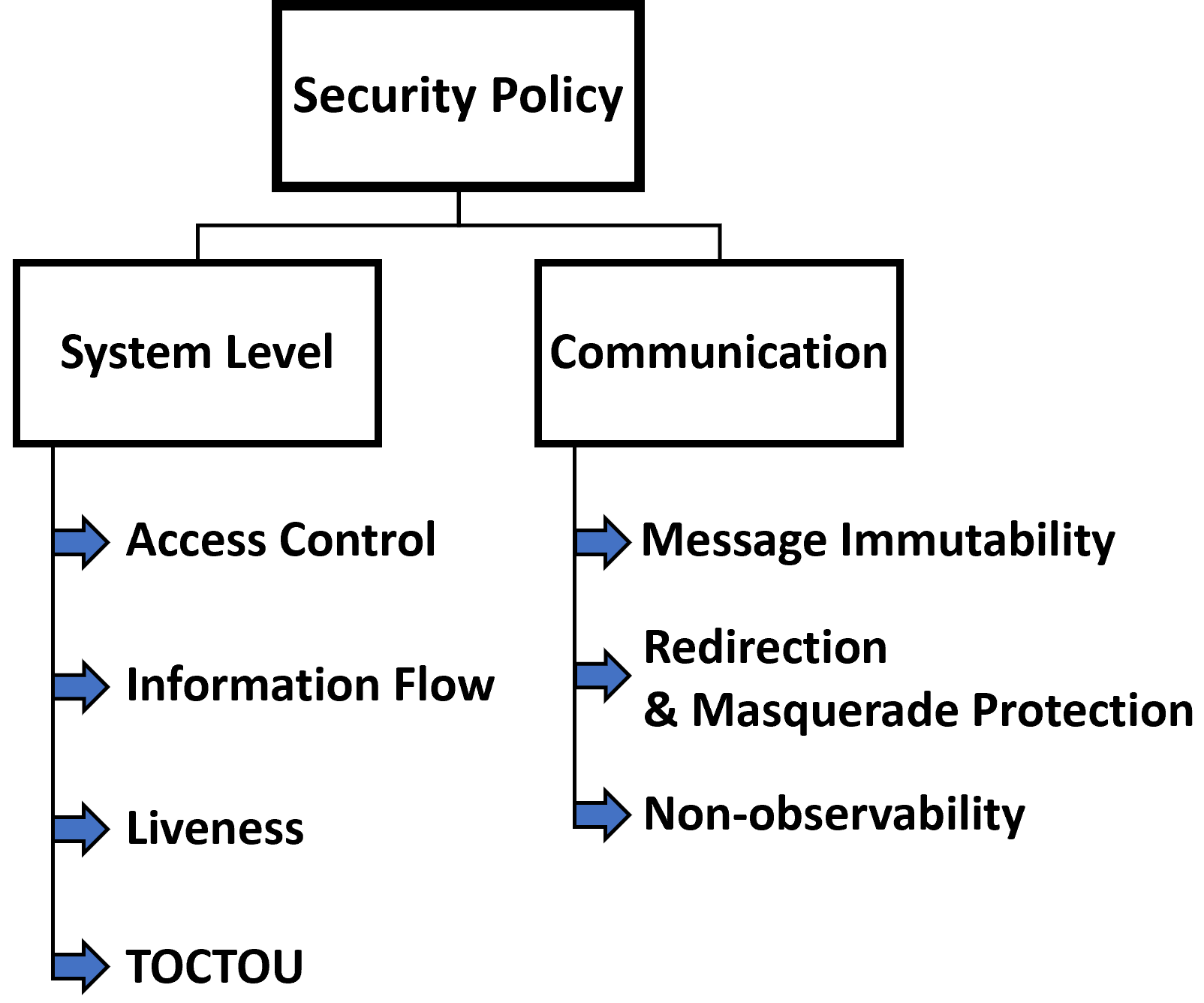}
\caption{Classification of Security Policies for Bus-based SoC in Current Practices}
\label{fig:policy_classification}
\end{figure}

The following taxonomy of security policy classes is presented in \cite{b16},\cite{b29} that, although not exhaustive, serve to demonstrate the range and variety of policies utilized in practice. Fig. \ref{fig:policy_classification} depicts a visual representation of the security policy classification recognized in practice.

\begin{itemize}
    \item Access Control: \\
    Access control policies represent the predominant class of security policies in SoC designs, governing the access permissions granted to different agents within the SoC at various stages of execution. An agent $A$, in this context, refers to a component that could be either software or hardware and present in any constituent IP within the SoC design. These access control policies serve as the fundamental ground for various additional policies, including integrity protection, information flow control, secure boot mechanisms, etc. 
    \item Information Flow: \\
    Information flow policies are employed to restrict the indirect inference of values of any secure asset. The values can be deduced by observing or snooping on intermediate computations or communications within IPs even when direct access is disabled. These policies restrict the flow of information between different components. In most cases, this class of policies is developed by combining access control policies with required auxiliary constraints. This ensures that sensitive information remains protected and inaccessible to unauthorized entities throughout the system. 
    \item Liveness: \\
    Liveness policies are designed to ensure that the implementation of protection mechanisms does not compromise the functionality of the system. This class of policies ensures that the system operates smoothly without facing any starvation during its execution. A typical liveness policy incorporates the response (grant or reject) mechanism following a request made by an IP for a resource. Violating such a policy can result in a livelock or deadlock situation, which can compromise the availability requirements of the system. By adhering to liveness policies, the system maintains its ability to perform its intended functions smoothly and continuously. 
    
    \item Time-of-Check vs. Time-of-Use (TOCTOU): \\
    TOCTOU policies aim to prevent the bypassing or spoofing of authentication mechanisms implemented for access control. These policies ensure that the authenticated agent is the one who is indeed accessing the asset for which it has been granted access. A crucial example of a Time-of-Check vs. Time-of-Use requirement arises in the context of firmware updates. In this scenario, the policy requires that the installed firmware during the updation process matches the previously authenticated firmware as mandated by the security engine. By enforcing TOCTOU policies, the system safeguards against unauthorized or tampered firmware, maintaining the integrity and security of the system's critical components. 
\end{itemize}

The subsequent categories provide a glimpse into the diverse requirements that need to be considered for on-chip communications in an SoC.
\begin{itemize}
    \item Message immutability: When an IP sends a message to another IP, then it is crucial that the message received by the receiver IP is precisely identical to the message sent by the sender IP.
    \item Redirection and masquerade prevention: When an IP transmits a message to another IP, then it is essential that the message is successfully delivered to the receiver IP without any compromise. It should be infeasible for a potentially rogue IP to masquerade as the actual receiver IP or redirect to another IP other than the intended IP.
    \item Non-observability: During transit, it is compulsory that a private message sent from one IP to another remains inaccessible to any other IP.
\end{itemize}

In many instances, security policies in SoC designs prioritize integration aspects over individual IPs. This approach means that the threat model mostly considers external attacks through software or the SoC interface with the system. However, the increasing adoption of 3rd party IPs offers a plug-and-play approach, allowing SoC design houses to meet tight time-to-market requirements using existing resources. As this trend continues, along with the static trust verification of IP designs, there arises a need to reevaluate and thoroughly analyze the trust model and assumptions underlying the current implementation of security policies in modern SoC designs.

In this work, we are focusing on enforcing different System Level Policies in any bus-based SoC. 

\subsection{Trust Model}

The security of System-on-Chip (SoC) designs is contingent upon the trust assumptions made regarding the design components and entities involved in the supply chain. In the context of this paper, the trust model assumes an SoC integration house to be trusted which acquires third-party intellectual properties (3PIPs) from multiple vendors, each with varying levels of trustworthiness. In accordance with this trust model, the IPs themselves may not be deemed trustworthy; however, the bus wrappers and communication between the IP-level bus wrapper and the Security Policy Module are considered to be trusted. From the perspective of SoC integration, the inherent trust model assumes that IPs are acquired from the supply chain at the global level, while the specified components must be carefully planned and designed by the SoC integration house. IPs in modern SoCs are regularly categorized based on different levels of trust. Security Policies are implemented to confirm that untrusted entities do not adversely affect the secure assets or have the capability to endanger other system components.

\textit{Assumption 1:} The currently proposed approaches assume the test and debug wrappers of the IPs to be trustworthy. In practice, the wrappers are designed in such a way that they are integrated into the SoC designs as infrastructure IPs to enable the debugging and testing tasks. Therefore, these test/debug wrappers have more stringent validation requirements than the conventional functional IPs in the SoC design. The wrappers are configured to monitor security-critical events and enforce the corresponding security policies.

\textit{Assumption 2:} To achieve a secure architecture, designating certain components as trusted is crucial. If every component is susceptible to compromise, ensuring the overall system's resilience becomes extremely challenging. SoC trust models aim to protect assets from malicious activities of entities which include IP vendors, SoC integration houses, foundries, test and validation facilities, etc. The underlying trust model assumes that the design components, such as Security Policy Module, Bus-level Wrappers, etc., specifically introduced for enforcing security policies are trusted, while individual IPs may be untrusted. 

\begin{table*}[!ht] 
\caption{Comparison with Existing Solutions \& Scope of the Current Work}
\label{tab:comparison}
 \begin{tabularx}{\textwidth}{X*{8}{>{\centering\arraybackslash}X}} 
 \hline \hline 
 Proposed Solutions & Working with Bus-Level SoC? & Centralized Policy Engine? &
 IP-Level Policies? & Support for Syn/Non-Syn Constructs? & Policies in Formal Representation? & Generalized Policies for Diverse Needs? & Automated Flow?
 \\ 
 \hline
 IIPS \cite{b17} & \cmark & \textcolor{red}{\xmark} & \cmark & \textcolor{red}{\xmark} & \textcolor{red}{\xmark} & \textcolor{red}{\xmark} & \textcolor{red}{\xmark} \\
 E-IIPS \cite{b12} & \cmark & \cmark & \cmark & \textcolor{red}{\xmark} & \textcolor{red}{\xmark} & \textcolor{red}{\xmark} & \textcolor{red}{\xmark} \\
 RSPE \cite{b26} & \cmark & \cmark & \cmark & \textcolor{red}{\xmark} & \cmark & \textcolor{red}{\xmark} & \textcolor{red}{\xmark} \\
 MSIPS \cite{b23} & \cmark & \cmark & \cmark & \textcolor{red}{\xmark} & \textcolor{red}{\xmark} & \cmark & \textcolor{red}{\xmark} \\
 DiSPEL* & \cmark & \cmark & \cmark & \cmark & \cmark & \cmark & \cmark \\
 \hline
\end{tabularx} 
 \footnotesize{ \\ * current work}
\end{table*}

\subsection{Threat Model}

To ensure the effective protection of assets, designers must establish security policies outlining protection requirements while also understanding the capabilities of potential adversaries. The effectiveness of security mechanisms in SoC designs relies on the realism of the adversary model considered during the development of protection strategies. It is essential to acknowledge that adversaries can vary depending on the specific asset being safeguarded. For example, DRM keys might consider end users as potential adversaries, while system manufacturers could be seen as potential threats to the secrecy of sensitive information. The process of defining and categorizing potential adversaries involves careful consideration and creativity while considering the factors like the physical access adversaries have to the system and their ability to observe, control, modify, or reverse-engineer components. By taking these factors into account, designers can develop more targeted and robust security measures to address the specific threats faced by the system. This approach enhances the overall resilience and reliability of the SoC design. In this work, we considered the following threat models in the context of securing assets in an SoC:

\begin{itemize}
    \item Access Control Violation \\
    This threat model involves an adversary attempting to gain unauthorized access to the SoC, bypassing security measures or exploiting vulnerabilities to gain control over the system or access sensitive information.

    \item Denial of Service (DoS)\\
    Adversaries in this threat model aim to disrupt or degrade the functionality of a bus-based SoC, rendering it unavailable or causing it to perform poorly. This can be achieved by flooding the system with excessive requests, exploiting resource limitations, introducing faults, etc.

    \item Information Leakage\\
    This threat model focuses on adversaries attempting to extract sensitive information from the SoC, such as encryption keys, confidential data etc., propagated or leaked through any observable point such as primary outputs, DFT, or DFD infrastructures.

    \item Side-Channel Leakage\\
    This threat model relies on the information obtained through side-channel leakage that allows an attacker to infer sensitive information by analyzing timing, power consumption, or electromagnetic emanation from the system. Such vulnerabilities can potentially expose critical information, making them a significant concern in SoC security.

    \item Reverse Engineering\\
    This threat model involves adversaries attempting to reverse engineer the bus-based SoC to understand its internal workings, identify vulnerabilities, or extract proprietary information.

    \item Tampering\\
    In this threat model, the attacker aims to disrupt or modify the connectivity service in a bus-based SoC, either physically or through remote means, to manipulate its behavior, compromise its integrity, or extract sensitive data.
\end{itemize}
It is important to note that extensive research has been conducted on specific attacks within these categories, but a detailed exploration of each model and attack is beyond the scope of this paper.

\subsection{Previous Work}

The early research during the 80s and 90s on security policies in a computing system was primarily focused on information security \cite{b1} and developing a framework for analyzing the information flow and defining access control (\cite{b2},\cite{b3}). However, the process of identifying and defining the policies was not found to be scalable for complex designs. Some researchers focused on developing languages to define security policies using formal representation. X. Li et al. \cite{b4} proposed a typed language called SAPPER which provides a synthesis framework for certain security policies. Some designers opted for other powerful languages, such as Property Specification Language or PSL (\cite{b15}), for defining and implementing security policies. Several researchers proposed another direction by representing security policies mapped to a specific class of Security Property. These properties can be tested statically using formal tools such as model checking \cite{b5}, equivalence checking \cite{b6}, and information flow tracking techniques (\cite{b7},\cite{b8},\cite{b10},\cite{b22}) for detecting information leakage, especially in cryptographic designs that cause confidentiality and integrity property violation. A technique named Gate-Level Information Flow Tracking (GLIFT) has been introduced to detect and measure illegal flows of a corrupted value at the Boolean level (\cite{b9}). Jin et al. \cite{b13,b14} proposed an analog on-chip neural network capable of detecting suspicious activity based on sensor measurements. Checking information flow helps to detect information leakage, which may infer the existence of Hardware Trojans \cite{b11}. 

There have been several works focused on developing security properties in the context of security verification and validation. These properties facilitate the verification of specific assumptions, requirements, conditions and the desired behaviors within a design. They formally describe the expected behaviors of the design concerning a particular security vulnerability. Evaluating the coverage of security properties can be used as a measure to assess the overall security of a SoC design. Authors in \cite{b21} proposed an automated property generation for information flow properties for hardware designs. Farzana et al. \cite{b19} created a database of security properties and generated equivalent assertions to verify them using appropriate tool sets (\cite{b20}) for identifying vulnerabilities in a design.

Ray et al. \cite{b16} discussed about incorporating security policies becoming an area of significant research activities with the advancements of modern SoC designs involving third-party IPs. The research directed towards incorporating a dedicated security IP for providing system-level protection for SoCs with untrusted IPs emerged as one of the viable solutions against diverse attacks. Wang et al. \cite{b17} proposed an infrastructure IP (IIPS) for SoC security architecture but limited to preventing low-level hardware security vulnerabilities. Basak et al. (\cite{b12,b18}) extended the IIPS framework and presented a microcontroller-based framework for implementing certain classes of security policies. Nath et al. (\cite{b26}) proposed Reconfigurable Security Policy Engine (RSPE), which can act as a smart security wrapper interfaced with the on-chip Design-for-Debug (DfD) interface for monitoring security-critical events and incorporating necessary mitigatory actions. Huang et al. \cite{b23} proposed an architecture called MSIPS that involves dedicated security IPs for enforcing security policies for protecting against H/W Trojan attacks, IP theft attacks, and some behavioral threats. 

Existing mechanisms typically address threats occurring individually during runtime at either the architecture level or IP core level and incorporate dedicated security IPs to withstand a common set of security threats. However, the current methodologies lack the flexibility to represent different security requirements and define actionable specifications using a high-level formal representation for any bus-based design. Our proposed architecture is capable of generating a synthesizable centralized module with minimal overhead for enforcing security requirements through policies from given user specifications.

\section{Methodology}

The primary goal is to provide systematic protection to secure assets against diverse attacks to uphold confidentiality, integrity, and availability (CIA) of any standard SoC design through appropriate security policies using an automatic tool flow. The proposed flow includes both manual and automation steps along with formal analysis while considering the design constraints. In the first step, the assets to be protected are identified through a manual process by analyzing the SoC architecture and the technical details provided by the architect/designer for each of the IPs involved. Some of the most common assets present in any standard SoC design are: Cryptographic keys, Digital Rights Management (DRM) keys, Programmable feature/defeature bits, Unlocking keys for locked IPs, Stored watermarks of IPs, On-chip debug instrumentation, Sensitive user data, etc. Once the assets are identified, the next step is to define the security policies involving different IPs and secure assets. The security policies are represented in a 3-tuple format for better readability and interpretability by the automated tool in the later stage. The formal representation of a security policy is represented in a 3-tuple format as follows:

\begin{center}
    $ Policy \# : <predicate, timing, action> $
\end{center} 

\noindent where,\\
\textit{predicate}: indicates specific conditions based on the IP-internal observable signals or property of the interconnection fabric and expressed as a Boolean function of multiple observable signals.\\
\textit{timing}: indicates either an operating mode or timing information with respect to the global clock.\\
\textit{action}: defines the step to be taken when the joint condition evaluates to be true achieved by asserting or de-asserting signals or performing specific checks on a set of variables.

\begin{figure}[!ht]
\centering
\includegraphics[scale=0.6]{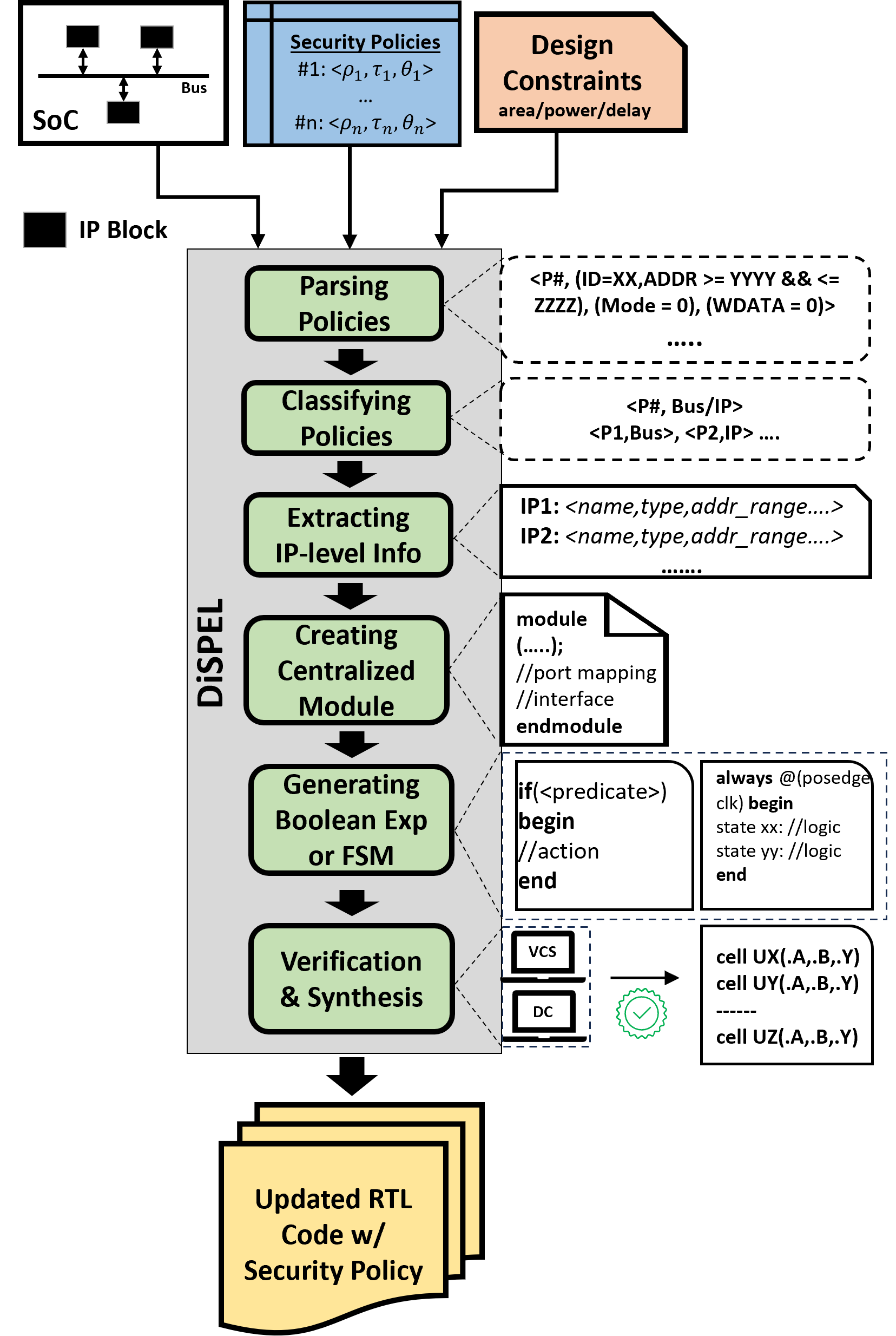}
\caption{Major Steps in DiSPEL Tool Workflow for Enforcement of Security Policies for a Bus-based SoC}
\label{fig:tool_workflow}
\end{figure}

The process of identifying and representing the relevant security policies using the above-mentioned format is a manual task done by the designer/architect/developer. The remaining steps in the proposed flow are automated, which can enforce the defined security policies in the form of SystemVerilog logic statements in a distributed manner across the bus interconnect between master and slave IPs. The automated security policy synthesizer, a.k.a DiSPEL, creates a wrapper on top of the SoC bus interconnect and controls the signals between master IPs and bus-interconnect and between bus-interconnect and slave IPs in both directions. The tool either passes the same value or alters the value of the signals based on the defined action behavior for different predicate and timing conditions of each security policy. The existing complexity in security policy analysis and modification arises from the fact that the policies are distributed across various IPs within the SoC. Hence, the DiSPEL tool in our proposed architecture works as an independently plugged centralized wrapper module on top of the SoC bus interconnect without altering the existing SoC design, thus alleviating the complexity. The workflow of the proposed framework has been depicted in Fig. \ref{fig:tool_workflow}.

\begin{figure}[!ht]
\centering
\includegraphics[scale=0.4]{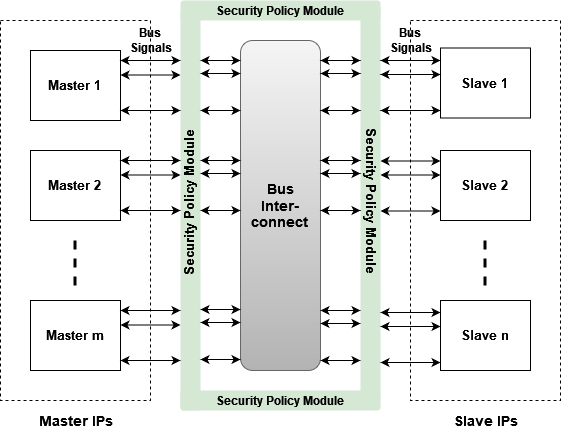}
\caption{Placement of Security Policy Module between IPs and Bus interconnect in any Bus-based SoC}
\label{fig:sec_pol_wrapper}
\end{figure}

The automated tool, DiSPEL, operates through six primary steps, outlined as follows:

\begin{enumerate}
    \item Parsing Security Policies\\
    In the initial step, the automated tool parses the identified relevant security policies represented using a 3-tuple format in a .json file. The tool employs regular expression-based parsing to detect the predicate, timing, and action fields for each policy and generate respective errors if there are any inconsistencies in the representation. The predicate, timing, and actions are stored individually in a data structure for each policy. For optional fields like timing/mode, the default value is assigned. 
    
     \item Classifying Policies\\
     Once the tool parses each policy in a 3-tuple format, it identifies whether it requires bus or IP-level enforcement from the given specifications and marks them accordingly. The automated tool is also capable of classifying each policy as either synthesizable or non-synthesizable. If the policy is marked as synthesizable, then the tool proceeds with the next action steps for enforcement of those policies in accordance with its workflow. For non-synthesizable policies, the tool generates SystemVerilog Assertions for verification purposes at the specified IP or at the bus level. Additionally, the tool can handle requirements that involve implementing FSMs for sequential events and complex clock cycle requirements. To enhance user convenience when defining policies in tuple format, we have included support for several common keywords (refer to Table \ref{tab:keywords}) that are translated to equivalent bus signal names by the tool.

    \begin{table}[]
    \centering
    \caption{List of supported keywords by DiSPEL Tool}
    \label{tab:keywords}
    \begin{tabular}{ccc}
    \hline
    & \multicolumn{2}{c}{Eqv. Bus Signals$^*$} \\
    \cline{2-3}
    Keywords & AXI4    & Wishbone \\
    \hline
    read\_address, read address, raddress & S\_AXI\_ARADDR      &  wb\_adr\_i       \\
    write\_address, write address, waddress &  S\_AXI\_AWADDR    &  wb\_adr\_o         \\
    read\_data, read data, rdata &    S\_AXI\_RDATA  &  wb\_dat\_i         \\
    write\_data, write data, wdata &   S\_AXI\_WDATA    &   wb\_dat\_o        \\
    strobe, strb, wstrb & S\_AXI\_WSTRB & wb\_stb\_i \\
    write\_ready, write ready, wready &   S\_AXI\_WREADY    &  -         \\
    read\_ready, read ready, rready &   S\_AXI\_RREADY    &    -       \\
    address\_ready, address ready, arready &   S\_AXI\_AWREADY    &   -        \\
    address\_valid, address valid, arvalid &   S\_AXI\_AWVALID   &  -         \\
    write\_valid, write valid, wvalid &   S\_AXI\_WVALID    &  -         \\
    read\_valid, read valid, rvalid &   S\_AXI\_RVALID   &  -         \\
    \hline
    \end{tabular}
        \caption*{
              \footnotesize
              *\underline{Note:} The representation of Eqv. Bus Signals are based on the MIT-CEP SoC benchmark and might need to be altered for other benchmarks accordingly.  
          } 
    \end{table}
    
    \item Extracting IP-level Information\\
    This information about IP level configuration is crucial for the proper functioning of the automated tool. It includes implementation details such as the address range, communication protocols, etc., specific to each master and slave IP. This information must be curated by the designer/architect/developer and given as input to the automated tool, which enables the tool to accurately analyze and process the IP configurations. Then, it can effectively generate appropriate code for enforcing the security policies at the respective IPs based on the IP level configuration data. Each master and slave IP must be provided with the these information: IP name, IP type (Master/Slave), Address range, Base/Starting/Ending Addresses, Starting \& Ending Data Marker.

    There are additional configuration details like global clock name, reset name, number of IPs, underlying bus protocol, etc., should be included in the configuration file which is given as input to the automated tool.
    
    \item Creating a Centralized Security Module \\ 
    Once the master and slave IP configurations of the SoC are extracted, the automated tool proceeds to generate a module that serves as an intermediary between the master/slave IPs and the bus interconnect. This module is designed to control the signals between these two components (refer to Fig. \ref{fig:sec_pol_wrapper}). The tool generates SystemVerilog code to incorporate this centralized module, which includes an interface containing all the input and output signals required for both the master and slave IPs of the SoC. By incorporating the centralized module, the automated tool ensures that the security requirements are met by enforcing the relevant security policies. The module acts as a security enforcement mechanism, monitoring and controlling the data flow between the IP interfaces and the bus interconnect to maintain a secure system operation. It helps protect sensitive information, maintain the confidentiality of data, and prevent unauthorized access to critical resources.

    \item Representation using Boolean Expression or Finite State Machine (FSM)\\
    The automated tool analyzes each security policy and evaluates it by identifying the conditions and actions specified for each policy. It then converts these policies into SystemVerilog conditional statements, utilizing boolean expressions to express the logic of the security requirements. For complex timing requirements or sequential behavior, the tool employs FSM-based state transitions and sets a flag when the conditions are met. The tool ensures that the correct IP modules are associated with the corresponding security policies by leveraging the given master/slave IP configurations. The tool follows the defined actions within each policy to assign the respective signal values for enforcing the security requirements to protect the secure assets in the SoC.
    
    \item Verification \& Synthesis\\
    The SystemVerilog code generated by the automated tool undergoes a thorough verification and validation process using commercial Electronic Design Automation (EDA) tools. These tools perform extensive checks and simulations to ensure the correctness and functionality of the generated code. During the verification phase, the code is subjected to various tests, including functional simulations, to validate its behavior against the intended design requirements. This helps identify and rectify any potential errors or inconsistencies in the code. Once the verification process is successfully completed, the next step involves mapping the SystemVerilog code to an equivalent gate-level netlist. This transformation is necessary for subsequent stages of the design process, such as physical implementation and synthesis. The tool also generates scripts for formal verification between the original and updated modules with security policies included.  
   
\end{enumerate}

The tool is  equipped with the ability to meet the specified design constraints in terms of area, power, or delay values using a feedback approach. In the initial synthesis, all the security policies are included as specified in the policy representation. If the constraints are satisfied, the tool retains each policy enforcement in the design. However, if the constraints cannot be met, the tool attempts to eliminate some of the policies with lower significance. To prioritize the policies, the tool employs a ranking mechanism based on the severity of potential exploitation. For example, a policy safeguarding a 32-bit secure asset for a crypto IP receives a higher score than a policy protecting a 1-bit value for an arithmetic calculation unit. Higher priority values are assigned to Crypto IPs, followed by Main Memory, HW Accelerators, and so on. The score is calculated for each policy as follows:

$score = f(Involved\_IP\_Type, No\_of\_Bits\_Protected, $\\ $ Type\_of\_Attack, Policy\_Type)$

The tool utilizes a feedback methodology to ensure adherence to the overhead constraints. If the constraints are not satisfied, DiSPEL uses an iterative approach to discard 10\% of policies based on their scores and reevaluates the overhead values. This iterative process continues until the design constraints are met. Table \ref{tab:score} provides a comprehensive overview of the scores assigned to various policies based on their severity levels. Each policy is identified by a unique number ({\em Policy\#}) and is associated with a specific {\em IP\_Type} within the SoC. The number of bits protected by each policy ({\em\#Bits\_Protected}) indicates the extent of sensitive information covered by that particular policy. The {\em Attack\_Type} include confidentiality (C), integrity (I), and availability (A) violations while {\em Policy\_Type} denotes if the policy is enforced at the bus level or at the IP level. The scores aid in identifying and prioritizing the most critical policies that require rigorous protection and validation to ensure the overall security of the system.

\begin{table}[!ht]
\centering
\caption{Representative Examples of Score Values after Normalization for Different Policies}
\label{tab:score}
\resizebox{\columnwidth}{!}{%
\begin{tabular}{c*{6}{c}r}
\hline
Policy\# & IP\_Type & \#Bits\_Protected & Attack\_Type & Policy\_Type & Score \\
\hline
P1 & Crypto & 128 & C,I,A & Bus & 19.2 \\ 
P2 & Hashing & 64 & C,I,A & Bus & 7.68 \\ 
P3 & Memory & 32 & C,A & IP & 0.96 \\
P4 & DSP & 16 & A & IP & 0.16 \\ 
\hline
\end{tabular}
}
\end{table}

\begin{algorithm}[]
\DontPrintSemicolon

  \KwInput{$\mathbb{F}$$_{soc}$, $\Phi$}
  \KwOutput{$\mathcal{F}$, $\mathcal{W^\prime}$}

   $module\_def,master\_block, slave\_block = \textbf{create\_security\_policy\_module($\mathbb{F}$$_{soc}$)}$ \\
   
   \For{each $\phi_i \in \Phi$}
   {
        $syn\_flag = false, fsm\_flag = false$ \\
        $\{\rho,\tau,\theta\} <= split(\phi_i)$ 
        $ip\_level = \textbf{classify\_policy}(\phi_i)$ \\
                   
            \If{$\textbf{is\_syn}(\rho)$} 
            {
                $syn\_flag = true$ \\
                
                \If{$\textbf{is\_sequential}(\rho) $ $or$ $\textbf{is\_clock\_cycles}(\tau)$} 
                {
                    $fsm\_flag = true$ 
                }
            } 
            \Else
            {
                $assertion$ = \textbf{create\_assertion}$(\rho,\tau,\mathbb{F}$$_{soc}$$)$ \\
                $\mathcal{W^\prime}_{id} = \mathcal{W}_{id}.append(id, assertion) $
            }
            \If{$syn\_flag$} 
            {
                \If{$\rho.\textbf{find}(keyowords)$}
                {
                    $\rho^\prime <= \textbf{replace\_keywords}(\rho)$ \\
                    $id, m\_flag, s\_flag = \textbf{extract\_id\_flag}(\rho^\prime)$ \\
                    $\sigma = split(\rho^\prime)$
                    
                }
                \Else
                {
                    $id, m\_flag, s\_flag = \textbf{extract\_id\_flag}(\rho)$ \\
                    $\sigma = split(\rho)$
                }

            \If{$fsm\_flag$}
            {
                 flag\_name = \textbf{create\_fsm($\sigma,\tau,\theta, \mathbb{F}$$_{soc}$)} \\ 
                 \If{$ip\_level$}
                 {
                        $\mathcal{W^\prime}_{id} =  \mathcal{W}_{id}.append(id,flag\_name)$
                 }
                 \Else
                 {
                     
                     $(m\_flag)?$ master\_block.$append(id,flag\_name)$ : slave\_block.$append(id,flag\_name)$
                 }
                   
            }
            \Else
            {
                 block = \textbf{create\_cond($id,\sigma,\tau,\theta$)} \\ 
                 \If{$ip\_level$}
                 {
                        $\mathcal{W^\prime}_{id} =  \mathcal{W}_{id}.append(block)$
                 }
                 \Else
                 {
                     $(m\_flag)?$ master\_block.$append(block)$ : slave\_block.$append(block)$
                 
                 }
                
            }  
        }
    }
   $\mathcal{F}.write$(\{$module\_def,master\_block,slave\_block$\}) \\
   \textit{return \textbf{$\mathcal{F},\mathcal{W^\prime}$}} 
\caption{Enforcement of Security Policies}
\end{algorithm}

Algorithm 1 describes the whole workflow of the DiSPEL automated tool in which we demonstrate the process of enforcing policies through a centralized policy module or appending to the corresponding bus-level wrapper for IP-level policies.
The inputs to the DiSPEL framework are described as follows: 
\begin{itemize}
    \item $\mathbb{F}$$_{soc}$: Denotes SoC configuration file, presented in JSON format, containing implementation details for each IP and the underlying bus protocol.
    \item $\Phi$ : Denotes the list of Security Policies such that, \\ $\forall i, \phi_i$ = $<predicate(\rho),timing(\tau),action(\theta)>$
\end{itemize}
The DiSPEL tool generates the centralized policy module $(\mathcal{F})$ and updated bus-level wrapper $(\mathcal{W^\prime})$ with security policies incorporated in them. 

Here are descriptions of some of the methods used in the algorithm:
\begin{itemize}
    \item \textit{create\_security\_policy\_module()}: This module creates the basic structure of the centralized module in SystemVerilog. The underlying test SoC configuration provided by the user in the configuration file is parsed to identify interface(s), bus signals, port definitions, etc.
    \item \textit{classify\_policy()}: This method returns \textit{true} if the security policy is an IP-level policy for any specific IP or \textit{false} if the policy is to be implemented through the centralized module. 
    \item \textit{is\_syn()}: This method returns \textit{true} if the security policy is found to be synthesizable. 
    \item \textit{is\_sequential()}: This method returns \textit{true} if the predicate consists of a sequence of events occurring in consecutive cycles or with a defined gap between cycles, necessitating the incorporation of an FSM.
    \item \textit{is\_clock\_cycles()}: This method returns \textit{true} if the security policy includes requirements related to counting clock cycles in the predicate or timing tuple.
    \item \textit{create\_assertion()}: This method generates SystemVerilog assertion from the given predicate and timing information for a policy marked as non-synthesizable. This enables the user to run functional verification to identify violations of security requirements. 
    \item \textit{replace\_keywords()}: This method replaces the allowed keywords with the respective bus signal names in the predicate tuple of a policy.  
    \item \textit{extract\_id\_flag()}: This method returns the identifier or ID number of the respective master or slave corresponding to the policy currently being parsed. 
    \item \textit{create\_fsm()}: This method generates a block representing an FSM with the necessary logic implementation according to the policy. It returns the name of the register that signifies the occurrence of the event when the condition in the predicate is satisfied. 
    \item \textit{create\_cond()}: This method generates a conditional block using \textit{if} statement and combining all the condition(s) specified in the predicate and timing tuple of the corresponding policy. It also generates the statement for assigning the respective bus signal values in accordance with the action tuple of the policy.
\end{itemize}

\section{Experimental Results}

\subsection{SoC Benchmark used for Testing \& Evaluation}
\begin{figure*}[!ht]
\centering
\includegraphics[scale=0.4]{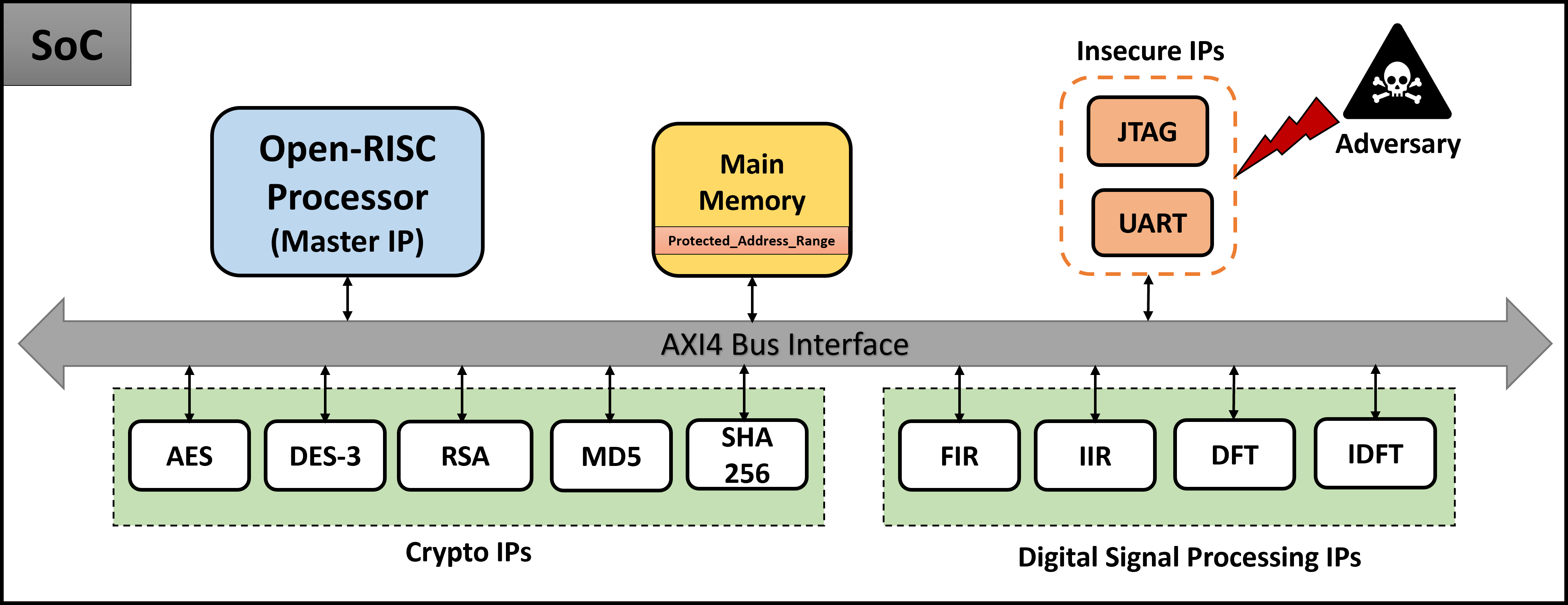}
\caption{A Representative Block Diagram of the SoC Benchmark with Master \& Slave IPs along with Insecure IPs connected through a Common Bus Interface}
\label{fig:soc_block_diagram}
\end{figure*}

This section provides an overview of the experimental setup and analysis of the obtained results to evaluate the performance of the proposed framework. To assess its effectiveness, we selected the CEP (Common Evaluation Platform), an open-source SoC benchmark developed by MIT, as our testing and evaluation platform. The CEP benchmark comprises various components, including the Mor1kx OpenRISC processor, which acts as the Master IP, and a set of Slave IPs with diverse functionalities. These Slave IPs consist of several cryptographic modules, such as AES, DES3, RSA, MD5, and SHA-256, as well as digital signal-processing modules like DFT, IDFT, FIR, IIR, and a GPS module. Additionally, there are certain modules like JTAG, GPIO, UART, etc., which facilitate external communication with the user. However, it is important to note that these modules may serve as potential attack surfaces and are therefore considered untrusted IPs. Fig. \ref{fig:soc_block_diagram} illustrates a basic block diagram of the SoC benchmark used in our experiment. We have categorized the Crypto and DSP modules along with  the open-source RISC processor as the master IP and the potentially insecure IPs communicating through the AXI4 bus interconnect.  

\subsection{An Example Security Policy: Representation \& Enforcement}

\noindent To provide a comprehensive insight into the automated generation process of synthesizable SystemVerilog code with security policy we will illustrate one such example policy below.

\noindent Policy \#: The processor (Master) is NOT ALLOWED to write data in between a specific ADDRESS RANGE of the RAM (Slave) in the user mode. 

\noindent \textbf{Representation in 3-tuple format:}

\begin{lstlisting}[style=code]
policy.json
{
	"Policy#":
	{
		"predicate": "address >= 0x0001dfa4 and address <= 0x0001ffac",
		"timing": "mode = 0",
		"action": "reject"
	}
}
\end{lstlisting}

\noindent DiSPEL tool will parse the security policy as follows:
\begin{itemize}
    \item The tool will classify Each Policy\# as either synthesizable or non-synthesizable. 
    
    The following steps will be carried out only if the policy is marked as synthesizable. 
    \item The predicate will be translated as: \\
\begin{lstlisting}[style=code]
((<address_var> >= 0x0001dfa4) && (<address_var> <= 0x0001ffac)) 
\end{lstlisting}
       
    \item The timing "mode = 0" will be checked as : \\
        $(<mode\_var> == 0)$                 //defines user mode 
    \item The tool will generate the conditional statement in SystemVerilog to enforce the security policy as: \\
\begin{lstlisting}[style=code]
if ((master.aw_addr >= 0x0001dfa4) && (master.aw_addr <= 0x0001ffac)) && (reg_mode == 0))
begin
    //action 
end 
\end{lstlisting}

    \item Based on the action definition, the respective signals will be updated. For this policy, the signal for writing data would be updated as: \\
\begin{lstlisting}[style=code]
    master_out.wdata <= 0;  //action
\end{lstlisting}

\end{itemize}

\begin{figure*}[!h]
\centering
\includegraphics[scale=0.5]{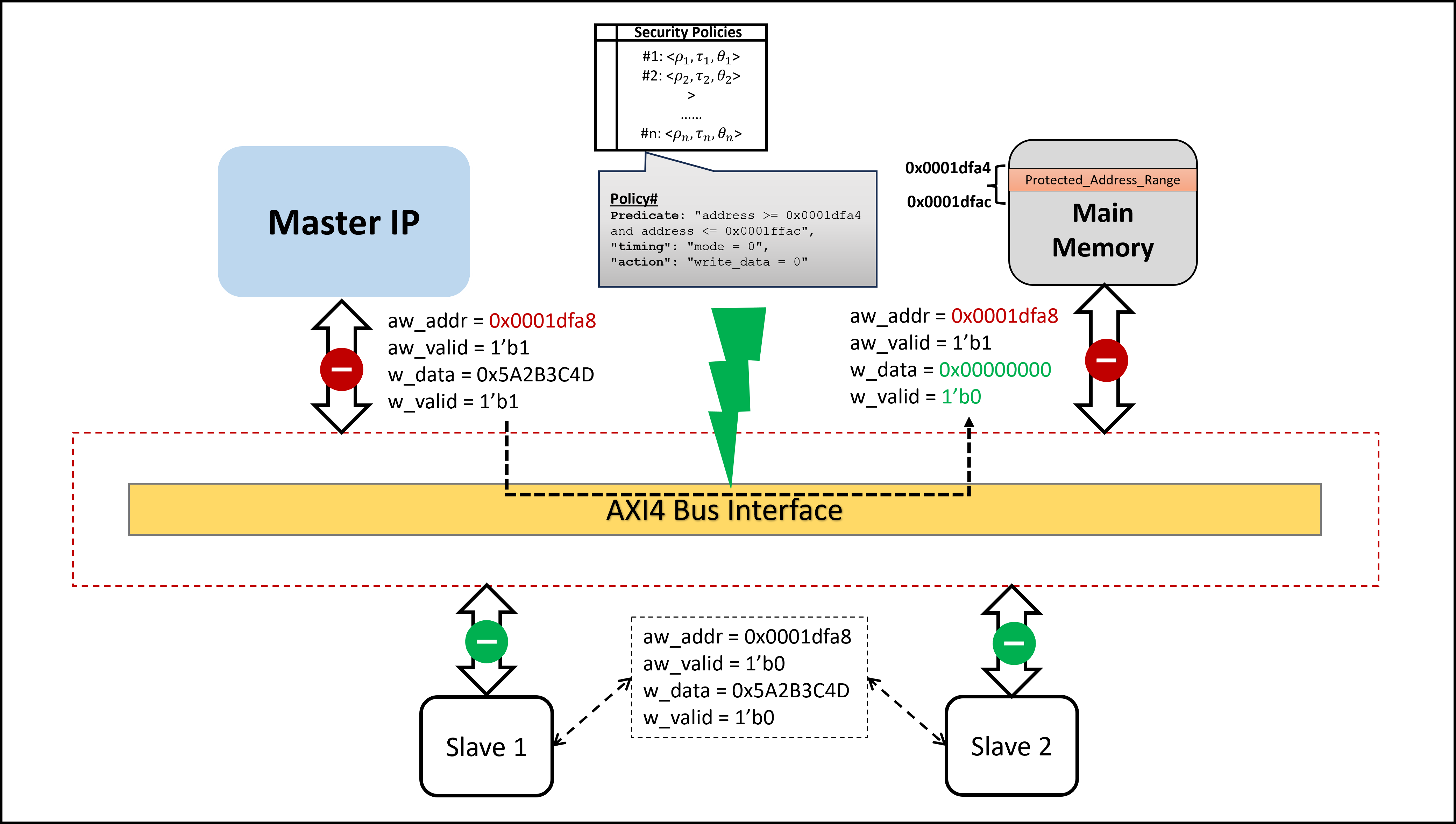}
\caption{Enforcement of an Example Security Policy by the Centralized Module involving multiple IPs in a Bus-based SoC}
\label{fig:working_soc_level_policy}
\end{figure*}

The enforcement of the security policy is illustrated in Fig. \ref{fig:working_soc_level_policy}. It demonstrates that when the address range falls within a protected region, the Master is prevented from writing $write\_data$ to the designated location. The diagram shows a snapshot of the communication between the Master IP (Processor) as the sender and the Slave IP (Memory) as the receiver, with other slave IPs connected through the bus interconnect. We have provided a description of the bus transactions for each IP, along with the enforcement of a relevant security policy at the bus level. A centralized module monitors all observable bus signals and updates their values if they do not comply with the security mandates. The bus transaction begins with the master sending the write data to a valid address value (0x0001dfa8) in the memory as indicated by the $aw\_valid$ signal. However, the underlying security policy restricts the writing of data between the address range starting at 0x0001dfa4 and ending at 0x0001ffac in memory as it contains sensitive data. Hence, the centralized policy module modifies the current bus transaction by updating the $w\_data$ to 0x0000000 and setting the $w\_valid$ to 0, which will restrict the writing of data in memory. The IPs involved in the bus transaction with the updation of values in the read/write channel are marked with red signs, while the slave IPs not involved in the transaction and have no updation to the read/write channel are marked as green. 

Table \ref{table:example_policy} represents a few example security policies applicable for either master or slave or both IPs and the policy representation in 3-tuple format with corresponding SystemVerilog construct for enforcement of each policy. 

\begin{table*}[ht]
\centering
\caption{Example Security Policies with their respective Policy Representation and SystemVerilog Constructs }
\label{table:example_policy}
\resizebox{\textwidth}{!}{%
\begin{tabular}{|l|c|l|l|}
\hline
\multicolumn{1}{|c|}{\textbf{Security   Policy}} &
  \textbf{IPs Involved} &
  \multicolumn{1}{c|}{\textbf{Policy Representation in 3-tuple   format}} &
  \multicolumn{1}{c|}{\textbf{SystemVerilog Constructs}} \\ \hline
\begin{tabular}[c]{@{}l@{}}Policy\#1: The Master must restrict the write data \\ operation for some specific addresses/address \\ range of the Slave.\end{tabular} &
  \begin{tabular}[c]{@{}c@{}}Processor (Master)\\Memory/AES/DES3/SHA-256 etc. (Slave)\end{tabular} &
  \begin{tabular}[c]{@{}l@{}}predicate : (write\_address \textgreater{}=   9300000c \\ and write\_address \textless{}= 93000010)\\      timing : "mode = 0"\\      action : "write\_data = 0"\end{tabular} &
  \begin{tabular}[c]{@{}l@{}}if(master{[}i{]}.aw\_addr \textgreater{}=   32'h9300000c \&\& \\ master{[}i{]}.aw\_addr \textless{}= 32'h93000010)\\      begin\\      master\_spw{[}i{]}.w\_data  = 0 ;\\      end\end{tabular} \\ \hline
\begin{tabular}[c]{@{}l@{}}Policy\#2: The Master is not allowed to read data \\ from a specific address range of the Slave.\end{tabular} &
  \begin{tabular}[c]{@{}c@{}}Processor (Master)\\      Memory/AES/DES3/SHA-256 etc. (Slave)\end{tabular} &
  \begin{tabular}[c]{@{}l@{}}predicate : (slave\_no = 1),   (read\_address \\ \textgreater{}= 93000004 and read\_address \textless{}= 93000008)\\      timing : "mode = 0"\\      action : "rdata = 0"\end{tabular} &
  \begin{tabular}[c]{@{}l@{}}if ((i==1) \&\&   (slave\_spw{[}i{]}.aw\_addr \textgreater{}= \\ 0x93000004 \&\& slave\_spw{[}i{]}.aw\_addr \textless{}=   \\ 0x93000008))\\      				begin\\      					slave{[}i{]}.r\_data = 0;\\      				end\end{tabular} \\ \hline
\begin{tabular}[c]{@{}l@{}}Policy\#3: The untrusted Slave(s) are not allowed \\ to access the bus when Secret Keys are \\ being sent to Slave Crypto Module.\end{tabular} & 
\begin{tabular}[c]{@{}c@{}}AES/DES3/RSA etc. (Trusted Slave) \\ JTAG/UART etc. (Untrusted Slave) \end{tabular} 
   &
  \begin{tabular}[c]{@{}l@{}}predicate : (slave\_no = 4 or 5),   (write\_address \\ \textgreater{}= 93000014 and write\_address \textless{}= 93000028)\\      timing : "mode = 0"\\      action : "wdata = 0"\end{tabular} &
  \begin{tabular}[c]{@{}l@{}}if ((!(i == 4 || i == 5)) \&\&   (slave\_spw{[}i{]}.aw\_addr \\ \textgreater{}= 0x93000014 \&\& slave\_spw{[}i{]}.aw\_addr \textless{}=   \\ 0x93000028))\\      begin\\      slave{[}i{]}.w\_data = 0;\\      end\end{tabular} \\ \hline
\begin{tabular}[c]{@{}l@{}}Policy\#4: If any particular Slave runs for more \\ than `n' number of cycles then discard\\ the result since the slave might be compromised.\end{tabular} &
  \begin{tabular}[c]{@{}c@{}}AES/DES3/RSA/SHA-256/MD5 etc. (Slave) \end{tabular} &
  \begin{tabular}[c]{@{}l@{}}predicate : (slave\_no = 2),   (clock\_cycles \textgreater 1000)\\      timing : "mode = 0"\\      action : "r\_data = 0"\end{tabular} &
  \begin{tabular}[c]{@{}l@{}}if ((i==2) \&\&   (flag==1))  //flag is a reg that   \\ //denotes if the condition is reached\\      begin\\      slave{[}i{]}.r\_data = 0;\\      end\\      //FSM to Count \# cycles while in Operation\\      …FSM States...\\      if(count \textgreater 1000) begin\\      flag \textless{}= 1;\\      curr\_state \textless{}= 1'b0;\\      end\end{tabular} \\ \hline
\end{tabular}%
}
\end{table*}

\subsection{Security Properties vs. Security Policies}

In the context of security verification, a security property serves as a statement to examine inferences, prerequisites, assumptions, and expected behaviors within a design. This statement can take the form of an assertion or a cover statement. Assertions are utilized to verify correct functioning and detect invalid events that may induce potential threats, while cover statements help in obtaining coverage information during the validation process of the design. Security properties can be categorized as immediate, checking specific scenarios at an instance of time, or concurrent, scrutinizing the behavior over a certain timeframe for the whole module. Concurrent assertions, being more potent, can express more complex events. Designers typically opt for assertion languages like PSL and SystemVerilog Assertions, which utilize logic representations at the temporal level like LTL and CTL (\cite{b24,b25}) to describe design behaviors. Various tools and methodologies are employed to verify the extracted security properties. Integrity properties, for instance, can be efficiently checked using model-checking tools, which excel in analyzing intricate systems to ensure correctness. In contrast, confidentiality properties are better assessed using information flow tracking and taint analysis techniques. These approaches can trace potential leaks of sensitive information to observable points within the design. Additionally, to examine properties related to secure design transformation, equivalence-checking tools are utilized, enabling verification of whether two designs exhibit identical behaviors.

Unlike non-synthesizable security properties, the proposed method incorporates DiSPEL tool that generates synthesizable RTL code in SystemVerilog. This enforces security policies for any bus-based SoC architecture. The synthesizable SystemVerilog code can be verified using simulation using any industry standard simulator, and it can also be verified on FPGA. The generated SystemVerilog code of the centralized module that enforces security policies inserted between the IPs and bus interconnect by the tool can be converted to a gate-level netlist using any standard EDA tools. Hence, the proposed architecture can be easily integrated into both ASIC and FPGA flow methodologies of any bus-based SoC design. 

Let us discuss an illustrative example to differentiate between security policy and security property.  \\
\textit{Security Requirement}: The secret key (K) should not be leaked (partially or fully) through Ciphertext (ct).

\noindent The following table (Table \ref{table:property_vs_policy}) demonstrates the above-mentioned security requirement using security property and in terms of security policy based on SystemVerilog conditional statement generated by DiSPEL tool. \\

\begin{table*}[t]
\centering
\caption{Enforcement of a specific security requirement through Security Property and Security Policy}
\label{table:property_vs_policy}
\resizebox{\textwidth}{!}{%
\begin{tabular}{|p{0.1\linewidth} | p{0.1\linewidth}| p{0.2\linewidth}| p{0.5\linewidth}| p{0.2\linewidth}|}
\hline
\multicolumn{1}{|c|}{Abstraction   Level} &
  \multicolumn{1}{c|}{IPs Involved} &
  \multicolumn{1}{c|}{Security Property} &
  \multicolumn{1}{c|}{Security Policy} \\ \hline
  \multicolumn{1}{|c|}{RTL} &
  AES (Slave IP) &
  \begin{tabular}[c]{@{}c@{}}no\_key\_leakage\_to\_output \\ := assert (ct != linear\_func(K))\end{tabular} &
  \begin{tabular}[c]{@{}l@{}}
  if(\$countones(ct{[}127:96{]} $^\wedge$ key{[}0{]})<8 ||   \$countones(ct{[}127:96{]} $^\wedge$ key{[}1{]})<8 \\ || \$countones(ct{[}127:96{]} $^\wedge$ key{[}2{]})<8  || \$countones(ct{[}127:96{]} $^\wedge$ key{[}3{]})<8 \\ ||   \$countones(ct{[}127:96{]} $^\wedge$ key{[}4{]})<8 || \$countones(ct{[}127:96{]} $^\wedge$ key{[}5{]})<8))  \\    wb\_dat\_o \textless{}= 32'b0;\\      else \\      wb\_dat\_o = ct{[}127:96{]};\\
  //repeat for ct[95:64], ct[63:32] \& ct[31:0] \\
  \end{tabular} \\ \hline
\end{tabular}%
}
\end{table*}

\subsection{Overhead Analysis} 

The representative overhead values for the IPs were obtained by synthesizing them at a 250nm standard cell library using Synopsys Design Compiler. Table \ref{tab:overhead_table_central} represents the overhead results in terms of area, delay, and power consumption for the centralized policy module implemented with different numbers of policies. Table \ref{tab:overhead_table_ip} demonstrates the extra incurred overhead for different IPs with different numbers of IP-level policies implemented. The results clearly demonstrate that the overheads are generally minimal under default synthesis settings without constraints. In certain cases, the overall overheads are even reduced after re-synthesizing with the implementation of the security policy module due to internal heuristic optimizations. Hence, we can conclude that our proposed methodology, which enforces security requirements through policies in a generic bus-based SoC design, imposes minimal overheads and is practically feasible to implement. 

\begin{table}[]
\centering
\caption{Centralized Module Implementation Results}
\label{tab:overhead_table_central}
\begin{tabular}{cccc}
\hline
& \multicolumn{3}{c}{\textbf{Synthesis Results}} \\
\cline{2-4}
\textbf{\#Policies} & \textbf{Area ($\mu m^2 $)}    & \textbf{Delay ($ns$)} & \textbf{Power ($mW$)} \\
\hline
10 & 69117.25      & 2.68    & 5.3295      \\
20 &  81799.39     &  2.63   &  7.4499     \\
25 &  71248.92     &  2.05   &  6.1466     \\
30 &  121181.24     &  2.05  &  11.29      \\

\hline
\end{tabular}
\end{table}

\begin{table}[h]
\centering
\caption{Overhead Analysis after Synthesis of Different IPs after Security Policy Enforcement}
\label{tab:overhead_table_ip}
\begin{tabular}{c|c|ccc}
\hline
\multirow{2}*{\textbf{IP}} & \multirow{2}*{\textbf{\#Policies}} & \multicolumn{3}{c}{\textbf{Synthesis Overheads (\%)}} \\
\cline{3-5}
 & & \textbf{Area}    & \textbf{Delay} & \textbf{Power} \\
\hline
AES & 10 & $0.09\uparrow$ & $-2.39\downarrow$ & $ 35.37\uparrow $\\ 
DES3 & 10 & $7.03\uparrow$ & $-15.21\downarrow$ & $27.22\uparrow $\\ 
SHA256 & 10 & $2.86\uparrow$ & $6.08\downarrow$ & $13.54\uparrow $\\ 
MD5 & 5 & $3.49\uparrow$ & $9.34\downarrow$ & $10.90\uparrow $\\ 
RSA & 5 & $0.05\uparrow$ & $-0.03\downarrow$ & $-15.25\uparrow $\\ 
\hline
\end{tabular}
\end{table}


\subsection{Waveform Analysis}
\begin{figure*}%
    \centering
    \subfloat[][]{\includegraphics[scale=0.6]{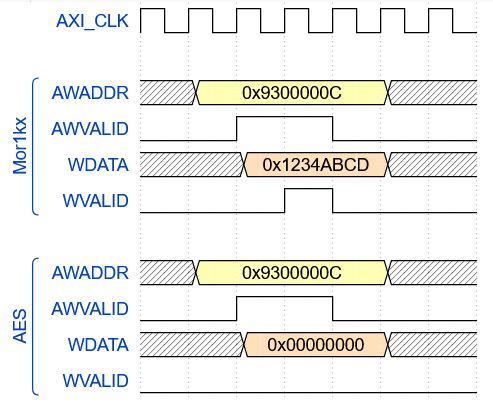}}
    \hfill
    \subfloat[][]{\includegraphics[scale=0.6]{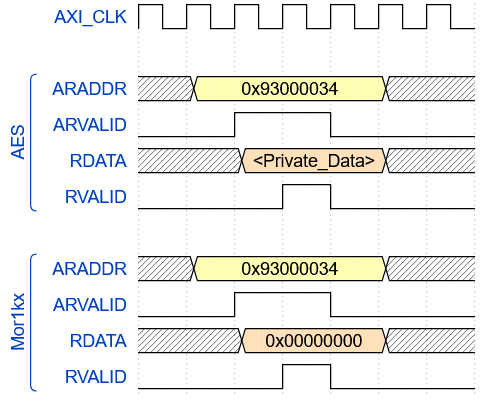}}
     \hfill
     \subfloat[][]{\includegraphics[scale=0.6]{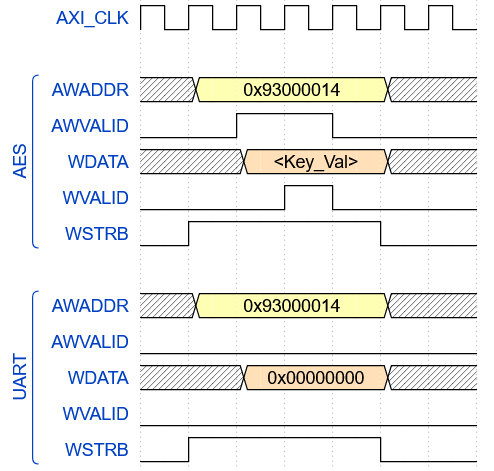}}
     \hfill
     \subfloat[][]{\includegraphics[scale=0.6]{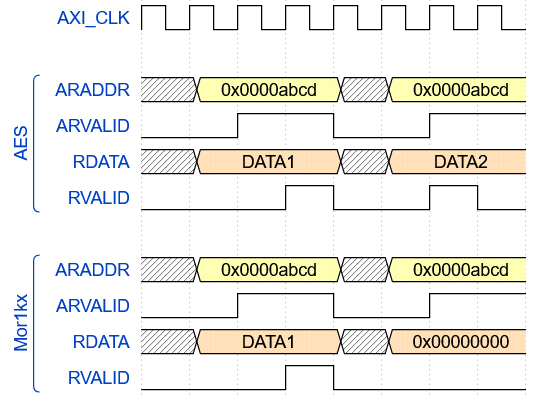}}
     \hfill
    \caption{Enforcement of Security Policies represented through timing diagrams: (a) Policy \#1 involving Mor1kx (Master) and AES IP (Slave), (b) Policy \#2 involving AES IP (Slave) and Mor1kx(Master), (c) Policy \#3 involving AES IP (Slave) and UART (Slave), (d) Policy \#4 involving AES IP (Slave) and Mor1kx(Master).}
    \label{fig:waveform}
\end{figure*}

The waveform analysis was a critical aspect of our experiment as it affirmed the effectiveness of enforcing a security policy that addresses the requirement of safeguarding private keys from untrusted IPs during transit. We generated the waveforms from the simulation using Synopsys VCS (Verilog Compiler Simulator) and employed a graphical user interface using DVE (Discovery Virtualization Environment) for viewing waveforms. For our experiment, we utilized Synopsys VCS Version: T-2022.06-SP2 and DVE Version: T-2022.06\_Full64 for analyzing the waveforms. We have illustrated the successful enforcement of the security policies explained in Table \ref{table:example_policy} through representative timing diagrams of related bus signals for the IPs involved in Fig. \ref{fig:waveform}. The demonstration clearly portrays that the private keys in the write channel of the AXI bus are only accessible to the intended AES IP but not exposed to untrusted IP. 

\subsection{Meeting Design Overhead Constraints}

\begin{figure}[h]
\centering
\includegraphics[scale=0.4]{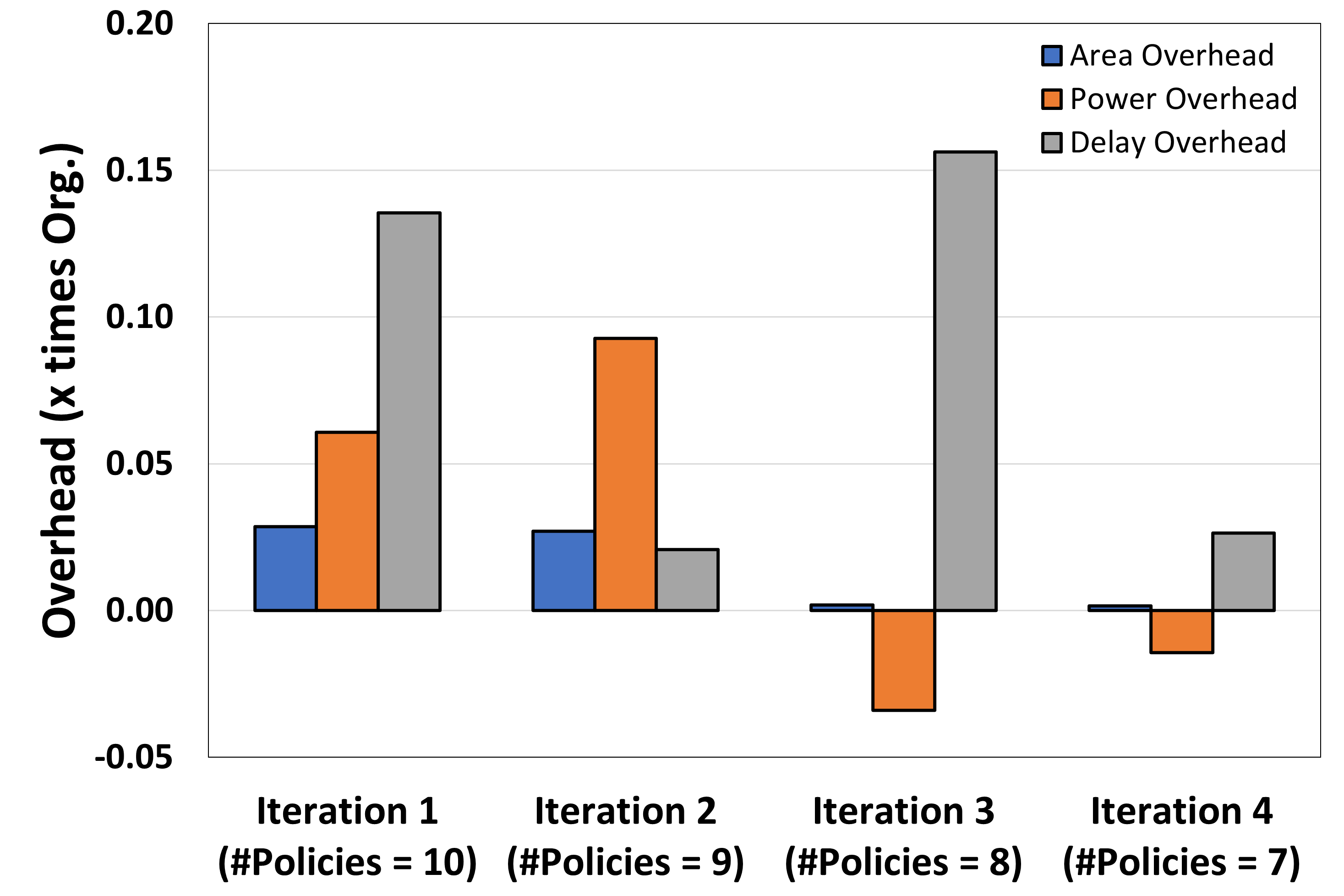}
\caption{Overhead Values represented as x times the Original Design and multiple iterations to meet the Overhead Constraints of SHA-256 Design}
\label{fig:sha_result}
\end{figure}

The DiSPEL Tool provides the flexibility to meet design constraints, considering area, power, and delay overhead values. The tool employs a ranking scheme based on severity to prioritize policies. After synthesizing the centralized module and the updated IP modules with the policies enforced, the tool utilizes a feedback-based approach to verify if the constraints are satisfied. If the constraints are not met, the tool adopts an iterative process. It discards the bottom 10\% of policies based on the score and performs re-synthesis to assess if the constraints are met in the subsequent iteration. This iterative process continues until the design constraints are successfully satisfied. Fig. \ref{fig:sha_result} depicts multiple iterations for the SHA-256 module from our test SoC benchmark. The initial constraints were set to less than 5\% allowed overhead for each area, power, and delay parameters. The designs are synthesized using 250nm standard cell library cells without any constraints. As shown in the figure, the initial synthesis resulted in violations of both power and delay overhead constraints. However, after discarding 10\% of the policies and performing re-synthesis in the subsequent iterations, all the overhead constraints were eventually met for the module. The effectiveness of DiSPEL tool was further tested by applying time constraints-based synthesis and experimenting with different IP modules in multiple iterations until the allowed constraints were met.

\subsection{Simulation \& Formal Verification Results}

The effectiveness of enforcing security policies can be demonstrated through simulation, while its correctness can be proved by formal verification. We have used Synopsys VCS (version: T-2022.06-SP2) for running the simulation tests for the overall SoC through bare metal code and unit-level simulation for each individual IP. The simulation results ensure the functional correctness of the design and allow for iterative design improvements and bug fixes if any violations are found. 

We opted for one of the most commonly used formal verification tools in the semiconductor industry, namely Cadence JasperGold v2020.12, for verifying the correctness of the policies. We represented the policies in the form of SystemVerilog Assertions and analyzed them through JasperGold coverage analysis. We performed comprehensive validation of all the policies to ensure their compliance with the specified security requirements. Table \ref{tab:simulation_formal_results} illustrates the simulation status and formal verification results for an example access control security policy involving address range validation for crypto IPs of the SoC.

\begin{table*}[!ht]
\centering
\caption{Simulation \& Verification Results after Security Policy Enforcement}
\label{tab:simulation_formal_results}
\resizebox{\textwidth}{!}{%
\begin{tabular}{|c|c|c|c|}
\hline
\textbf{IPs   Involved} & \textbf{SystemVerilog Assertion}                                                                                             & \textbf{Simulation   Status} & \textbf{Formal   Verification Status} \\ \hline
AES                     & wb\_adr\_i \textless{}=   32'h93000000 \&\& wb\_adr\_i \textgreater{}= 32'h9300FFFF \&\& wb\_dat\_o ==   32'h0 & Passed                       & Covered                               \\ \hline
DES3                    & wb\_adr\_i \textless{}=   32'h97000000 \&\& wb\_adr\_i \textgreater{}= 32'h9700FFFF \&\& wb\_dat\_o ==   32'h0 & Passed                       & Covered                               \\ \hline
MD5                     & wb\_adr\_i \textless{}=   32'h94000000 \&\& wb\_adr\_i \textgreater{}= 32'h9400FFFF \&\& wb\_dat\_o ==   32'h0 & Passed                       & Covered                               \\ \hline
RSA                     & wb\_adr\_i \textless{}=   32'h96000000 \&\& wb\_adr\_i \textgreater{}= 32'h9600FFFF \&\& wb\_dat\_o ==   32'h0 & Passed                       & Covered                               \\ \hline
SHA256                  & wb\_adr\_i \textless{}=   32'h95000000 \&\& wb\_adr\_i \textgreater{}= 32'h9500FFFF \&\& wb\_dat\_o ==   32'h0 & Passed                       & Covered                               \\ \hline
\end{tabular}%
}
\end{table*}

\section{Conclusion \& Future Work}

We have introduced a distributed automated tool flow that enables the implementation of diverse security policies in a system-on-chip (SoC). This framework simplifies the process of secure SoC design by providing a systematic approach for incorporating various security assets while potentially reducing design and hardware overhead. Additionally, it facilitates the process of defining security requirements and enforcing security policies which is often a major challenge in the SoC production cycle. The automated tool flow incorporates a centralized security policy module for bus-level policies and updates the bus-level wrapper for each IP block for IP-level policies. This framework is scalable and can accommodate a large number of IPs with different structural properties for any bus-based SoC. We have validated the functional correctness of the architecture through extensive simulations and also employed formal verification. We evaluated the hardware overhead in terms of area, power, and delay and the results are very promising as it incurs significantly less overhead for practical SoCs. Future work will accommodate more bus protocols with different SoC configurations and can be extended to NoC configurations.

\section*{Acknowledgment}
The work was supported by Air Force Research Laboratory (AFRL) under the STAMP grant.

\end{document}